\documentclass[reqno,11pt]{article}
\usepackage{amssymb, amsmath}
\usepackage{graphicx}
\usepackage{graphics}

\newcommand{\C}{\mathbb{C}}
\newcommand{\CP}{\mathbb{CP}}

\def\be{\begin{equation}}
\def\ee{\end{equation}}
\def\theequation{\thesection.\arabic{equation}}

\def\p{\partial}
\def\tw{\widetilde{W}}
\def\tz{\widetilde{Z}}

\usepackage{bm}
\usepackage{enumerate}

\def\theequation{\thesection.\arabic{equation}}

\newcounter{mnotecount}[section]

\renewcommand{\themnotecount}{\thesection.\arabic{mnotecount}}

\newcommand{\mnote}[1]%{}%
{\protect{\stepcounter{mnotecount}}$^{\mbox{\footnotesize
$%\!\!\!\!\!\!\,
\bullet$\themnotecount}}$ \marginpar{%\color{red}%
\raggedright\tiny\em
$\!\!\!\!\!\!\,\bullet$\themnotecount: #1} }

\def\theequation{\thesection.\arabic{equation}}
\numberwithin{equation}{section}

\begin{document}
\date{29 May 2024}
%%%%%%%%%%%%%%%%%%%%%%%%%%%%%%%%%%%%%%%%
\title{Twistor theory of the Chen--Teo gravitational instanton}
%%%%%%%%%%%%%%%%%%%%%%%%%%%%%%%%%%%%%%%%
\author{Maciej Dunajski\\
  Department of Applied Mathematics and Theoretical Physics\\ 
University of Cambridge\\ Wilberforce Road, Cambridge CB3 0WA, UK\\
{\tt m.dunajski@damtp.cam.ac.uk}\\ \\
and\\ \\
Paul Tod\\
The Mathematical Institute\\
Oxford University\\
Woodstock Road, Oxford OX2 6GG, UK.\\
{\tt tod@maths.ox.ac.uk}
}

%%%%%%%%%%%%%%%%%%%%%%%%%%%%%%%%%%%%%%%%%%%%%%%%%%%%%%%%%%
\maketitle
\begin{center}
{\em Dedicated to Nick Woodhouse on the occasion of his 75th birthday}
\end{center}
\begin{abstract}
Toric Ricci--flat metrics in dimension four correspond to certain holomorphic vector bundles over a
twistor space. We construct these bundles explicitly, by exhibiting and characterising their patching matrices, 
for the five--parameter family of Riemannian ALF metrics constructed by Chen and Teo. The Chen--Teo family contains
a two--parameter family of asymptotically flat gravitational instantons. The patching matrices for these 
instantons take a simple rational form.
\end{abstract}
\section{Introduction}
In \cite{CT2} Chen and Teo constructed a five--parameter family of Riemannian Ricci--flat four--manifolds with torus symmetry. In the limiting cases this family contains
the Riemannian Pleba\'nski--Demia\'nski solutions \cite{PD} and some multi--centred Gibbons--Hawking metrics 
\cite{GH}, as well as a new two--parameter family of asymptotically flat gravitational instantons \cite{CT1}
on the four--manifold $\CP^2{\setminus}S^1$.

It was shown by Louis Witten \cite{Witten}, and in a different form by Richard Ward \cite{W2}, that any Lorentzian Ricci flat toric metric in four--dimensions corresponds to a static and axi--symmetric solution of the anti--self--dual Yang Mills equations
(ASDYM) with gauge group $SL(2, \C)$ on  flat space--time. These field equations are integrable by the twistor
correspondence of Ward \cite{W1}, and their solutions
are encoded in rank--two holomorphic vector bundles over the twistor space $PT\equiv\CP^3\setminus\CP^1$ which are trivial on twistor lines. Any such bundle can be defined in terms 
a patching matrix, and the Ward correspondence
gives an algorithm for recovering the ASDYM connection,
and eventually the toric vacuum solution from such a matrix. This construction has primarily focused
on Lorentzian toric vacuum solutions.

The aim of this paper is to adapt this construction to  Riemannian signature, and use it to 
put the Chen--Teo metrics in the twistor framework. In the next section we shall recall
the Chen--Teo metrics. In \S\ref{section2} we shall
use the rod structure of these metrics to construct
the patching matrix for the Ward bundle. It is
a rational matrix with three simple poles, and 
asymptotics encoding conserved charges.
In \S\ref{section3} we discuss the reconstruction
of the Chen--Teo patching matrix from a general patching matrix with three simple poles. The Ward correspondence for the reduced ASDYM equations is summarised in the Appendix.

\subsection*{Acknowledgements} 
We are grateful to James Lucietti for explaing to us how the mass of the Chen--Teo instanton which we compute in \S\ref{subinst} relates to the mass calculated in \cite{Lucietti}.
\section{The Chen--Teo Ricci--flat metrics}
Consider a quartic
\be
\label{quartic}
f=f(\xi)=a_4\xi^4+a_3 \xi^3+a_2\xi^2+a_1\xi+a_0
\ee
with four real roots and set
\begin{eqnarray}
\label{FGH}
F&=&f(x)y^2-f(y)x^2\\
H&=& (\nu x+y)[(\nu x-y)(a_1-a_3xy)-2(1-\nu)(a_0-a_4 x^2y^2)]\nonumber\\
G&=& f(x)[(2\nu-1)a_4 y^4+2\nu a_3  y^3+a_0 \nu^2]-f(y)[\nu^2 a_4 x^4+2\nu a_1 x+(2\nu-1)a_0].\nonumber
\end{eqnarray}
Let $(x, y, \tau, \phi)$ be local coordinates on an open set of a  four--manifold $M$.
The family of metrics
\be
\label{CTmetrics}
g=\frac{kH}{(x-y)^3}\Big(\frac{dx^2}{f(x)}-\frac{dy^2}{f(y)}-\frac{f(x)f(y)}{kF} d\phi^2\Big)
+\frac{1}{FH(x-y)}(F d\tau+Gd\phi)^2
\ee
is Ricci--flat \cite{CT2} for any values of the constant real parameters $(k, \nu, a_0, \cdots, a_4)$. Moreover 
two out of the five  quartic coefficients  can be fixed by constant scalings of the parameters and coordinates.
Thus (\ref{CTmetrics}) is a five--dimensional family of toric metrics, where the torus symmetry is generated
by constant translations of the $(\tau, \phi)$ coordinates.   In \cite{AA} it was subsequently shown, that all metrics in the Chen--Teo family are
Hermitian,  or equivalently one--sided type $D$: the self--dual part of the Weyl tensor is of Petrov--Penrose type $D$. This Hermitian structure is non--K\"ahler, unless the parameters
are chosen so that the metrics are self--dual or anti--self--dual. It is however the case that all Chen--Teo metrics are conformal to K\"ahler \cite{DTKahler}. Further global and local properties of
(\ref{CTmetrics}) have been uncovered in \cite{Lucietti, BG, Araneda}.

Let $r_1< r_2< r_3< r_4$ be four real roots of the quartic (\ref{quartic}) assumed to exist. To avoid curvature 
singularities of the metric resulting from vanishing of $H$, Chen and Teo \cite{CT1, CT2} restrict the range of the coordinates $(x, y)$ to
a rectangle 
$
-\infty<r_1<y<r_2<x<r_3<r_4<0
$ (Figure 1)
\begin{center}
\includegraphics[scale=0.3,angle=0]{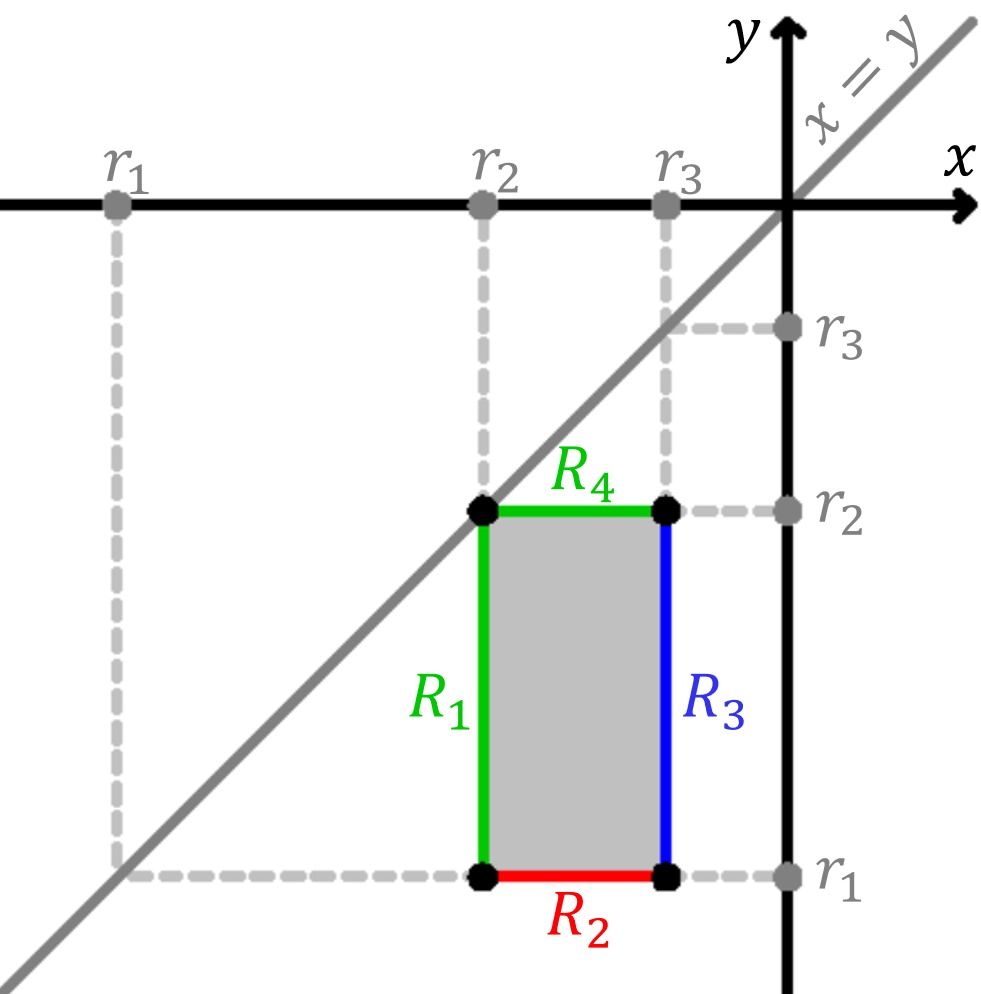}
\begin{center}
{{\bf Figure 1.} {Domain of $(x, y)$ dependence in the Chen--Teo metrics.}}
\end{center}
\end{center}
with the diagonal $x=y$ corresponding to the boundary at infinity of the asymptotic region. Further constraints
must then be imposed on these roots, as well as the parameter $\nu$, to avoid conical singularities and
to ensure asymptotic flatness (AF). The limiting cases $\nu=1$ and $\nu=-1$ correspond to the Pleba\'nski--Demia\'nski family \cite{PD}, and the ALE Gibbons--Hawking metrics \cite{GH} with three collinear centres respectively.

The AF condition is equivalent to setting the NUT parameter to zero,
and this is in turn equivalent to the vanishing of $n$ in (\ref{massnut}) below. In the rest of this paper,
with the exception of \S\ref{subinst} we shall
consider the Chen--Teo rectangular domain but will not make any restrictions on the parameters.
\section{The Yang equation and its patching matrix}
\label{section2}
The Chen--Teo metrics (\ref{CTmetrics}) can be put in the `two commuting Killing vectors form'
\be
\label{2kf}
g=\Omega^2(dr^2+dz^2) +J_{ij} d\phi^id\phi^j, \quad i, j=1, 2
\ee
where $\phi^i=(\phi, \tau)$, the Killing vectors are $K_i=\p/\p \phi^i$, and
the two by two matrix $J$ depends on the coordinates $(r, z)$ on the space
of orbits  of the $T^2$ action, with
\be
\label{rJ}
r^2=\mbox{det}(J).
\ee
Both $r$ and $z$ are harmonic on the orbit space and once $r$ is constructed from $J$, then $z$ is found, up to additive constant, as its harmonic conjugate. The equation (\ref{rJ}) implies that the matrix $J$ drops its rank
on the axis $r=0$, where rank$(J)$  is generically equal to one. The rank drops to zero
at points variously called {\em turning points} \cite{CT1}, {\em nuts} \cite{Metz} or {\em  nodes} : these are the isolated points corresponding
to the $z$ values where both Killing vectors vanish. The intervals connecting these points on the $r=0$ axes are  the {\em rods}, and there are two semi-infinite rods at the ends of the axis  corresponding to the edges
of the rectangle from Figure 1 which meet on the diagonal at $(r_2, r_2)$.
The Ricci--flat condition on $(M, g)$ gives the Yang equation
\be
\label{Yang}
r^{-1} \p_r (r J^{-1}\p_r J)+\p_z(J^{-1}\p_z J)=0
\ee
together with a linear equation relating $\Omega$ and $J$. 
\subsection{The patching matrix}
The Yang equation
(\ref{Yang}) also arises as a symmetry reduction of ASDYM by a translation and a
rotation. 
It is therefore integrable by the Ward correspondence \cite{W1}, and
the twistor data consists of a patching matrix for a bundle
on either the whole twistor space $PT=\CP^3\setminus\CP^1$ or its
reduction to the non--Hausdorff mini--twistor space \cite{W2, WM, FW} (see Appendix A). The algorithm
to construct this patching matrix consists of two steps \cite{WM}:
\begin{enumerate}
\item For any choice of $U(1)\subset T^2$ generated by the Killing vector
  $K$ the twist of $K$ is a gradient, i. e.
  \be
  \label{twist_pot}
    d\psi =* (K\wedge dK).
  \ee
  Then \cite{FW}
  \be
  \label{Jprime}
    J'=\frac{1}{V}
\begin{pmatrix}
1 & -\psi \\
-\psi & \psi^2-V^{2}
\end{pmatrix}, \quad \mbox{where}\quad V\equiv g(K, K)
\ee
is another solution to (\ref{Yang}) with\footnote{Note the relative sign difference between
$\psi^2$ and $V^{2}$ in the bottom right component of $J'$. In the Lorentzian case this relative sign is $+$,
and $\mbox{det}(J')=1$.} $\mbox{det}(J')=-1$.
\item Pick a rod on which $K$ is not identically zero. Then
  \be
	\label{Pmatrix}
  P=J'(0, z)
  \ee
  is regular on this rod, except for the poles at its end--points, and
  is the patching matrix for the holomorphic vector bundle
  over the reduced twistor space.
\end{enumerate}
We shall implement this for the Chen-Teo class (\ref{CTmetrics})
which contains the four rods  corresponding to the edges
of the rectangle on Figure 1, with two--semi infinite outer rods $R_1, R_4$.
To transform the metric  to the $(r, z)$ coordinates use 
\begin{eqnarray*}
  z&=&\frac{2a_4 x^2y^2+2a_2xy+2a_0+a_3xy(x+y)+a_1(x+y)}{2(x-y)^2}\\
       r&=& \frac{\sqrt{-f(x)f(y)}}{(x-y)^2},
 \end{eqnarray*}
and find the $z$--values at the turning points $(x, y)=(r_2, r_1), (r_3, r_1), (r_3, r_2)$ to be
\[
z_3=-a_4(r_1r_2+r_3r_4)/2, \quad
z_2=-a_4(r_1r_3+r_2r_4)/2, \quad
z_1=-a_4(r_1r_4+r_2r_3)/2.
\]
The rod structure is now encoded on the $z$--axis which contains four edges of the rectangle from Figure 1,
with the edges meeting at $(x, y)=(r_2, r_2)$ going out to $\pm \infty$ (Figure 2). 
\begin{center}
\includegraphics[scale=0.3,angle=0]{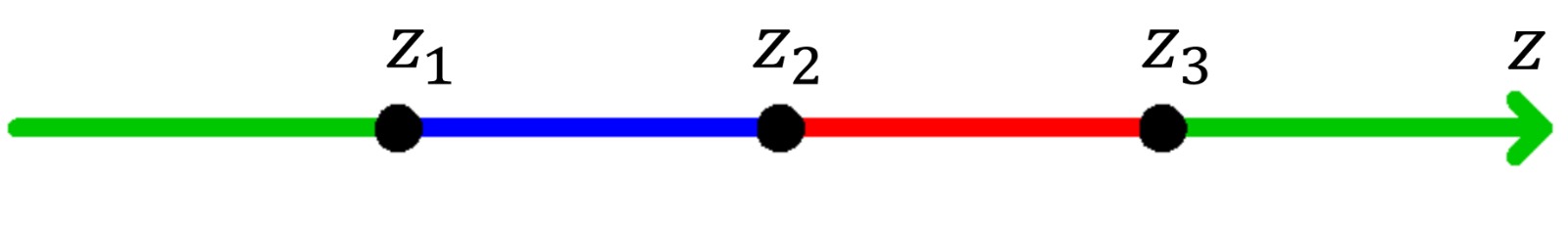}
\begin{center}
{{\bf Figure 2.} {The rod structure of the Chen--Teo family.}}
\end{center}
\end{center}
We shall focus on the semi--infinite rod, where $x=r_2$. The Killing vector 
$K=\sqrt{1-\nu^2}\p/\p\tau$ does not vanish on this rod, and
\be
\label{Weq}
V\equiv g(K, K) =\frac{(1-\nu^2)F}{(x-y)H}.
\ee
The corresponding twist potential $\psi$ in (\ref{twist_pot}) is defined up to  addition of a constant, and we shall choose this
constant so that $V^{-1}\psi$ tends to zero at the infinite end point $x=y=r_2$ ( or $r=0, z=\infty$) of the rod. This yields
\be
\label{psi_eq}
\psi=\frac{(y-x)(a_4 x^2y^2(1-\nu)^2
-a_3\nu xy(x+y)+a_1\nu (x+y)+a_0(1-\nu)^2)}{H}
%\frac{a_4[(1-2\nu)x^3y^2-x^2y^3]-a_3\nu x^3 %y+a_1\nu x^2+a_0[(2\nu-1)x+y]}{H} +\nu,
\ee
where $H$ is given by (\ref{FGH}). To transform $\psi$ and $V$ to the $(r, z)$ coordinates we shall  set
$x=r_2$ (this defines rod $R_1$), and solve for $y$ in terms of $z$ as
\[
y=\frac{2 r_2 z+a_4(r_1r_2r_3+r_1r_2r_4-r_1r_3r_4+r_3r_2r_4)}{2z+a_4(r_1r_2+r_3r_2+r_4r_2-{r_2}^2)}.
\]
Substitute this to (\ref{psi_eq}) and (\ref{Weq}) to find formulae for $\psi$ and $V$ which are rational functions
of $z$.  Substituting these formulae 
into (\ref{Jprime}) and (\ref{Pmatrix}) yields the twistor patching matrix
\be
\label{patch_final}
P(z)=\begin{pmatrix}
C_1/C & Q/C   \\
Q/C & C_2/C
\end{pmatrix},
\ee
where $Q$ is a quadratic and $C, C_1, C_2$ are monic cubics given by
\begin{eqnarray}
\label{coeffC}
C&=&(z-z_1)(z-z_2)(z-z_3)=z^3+c_2 z^2+c_1 z+c_0\nonumber\\
C_1&=& z^3+c_{12} z^2+c_{11} z+
c_{10}\nonumber\\
C_2&=&- z^3+c_{22} z^2+c_{21} z+
c_{20}\nonumber\\
Q&=& q_2 z^2+q_1 z+q_0.
\end{eqnarray}
The coefficients of $c_i, c_{1i}, c_{2i}, q_i, i=0, 1, 2$  of $C, C_1, C_2, Q$ are listed in Appendix B.

This patching matrix has three poles, but two of them do not belong to the rod we are considering. 
There is a known procedure \cite{F, FW} to pass through a turning point $z=z_3$ and construct the patching matrix
corresponding to the adjacent rod connecting $z_3$ to $z_2$. This can be repeated to construct patching  
matrices for all four rods starting from $P$ in (\ref{patch_final}). The asymptotics of $P$ near the semi--infinite end of the rod is
\be
\label{asymptoticP}
P\cong
\begin{pmatrix}
1 & 0   \\
0 & -1
\end{pmatrix}+
\frac{1}{z}\begin{pmatrix}
2m & 2n   \\
2n & 2m
\end{pmatrix}
+O(1/z^2),
\ee
where
\begin{eqnarray}
\label{massnut}
m&=&\frac{a_{4} \left(2 \nu  r_{2}^{3}+\left(\nu^{2}+1\right) \left(r_{1}+r_{3}+r_{4}\right) r_{2}^{2}-\left(r_{3}r_1+r_{4} r_{1}+r_{3} r_{4}\right) \left(\nu^{2}+1\right) r_{2}-2 \nu  r_{1} r_{3} r_{4}\right)}{2 \left(2 \nu^{2}-2\right) r_{2}}\\
n&=&\frac{q_2}{2}=
 \frac{\left(-2 \nu  \left(r_{1}+r_{3}+r_{4}\right) r_{2}^{2}+2 \nu  \left(r_{3}r_1+r_{4}r_{1}+r_{3} r_{4}\right) r_{2}+ \left(\nu^{2}+1\right)( r_4r_{3} r_{1} -{r_2}^3) \right) a_{4}}{\left(4 \nu^{2}-4\right) r_{2}}.
 \nonumber
\end{eqnarray}
James Lucietti has pointed out to us how $(m,  n)$ are related to the mass and NUT parameter which we shall now explain. To exhibit the ALF property of their metrics Chen and Teo define $(\tilde{\tau}, \tilde{\phi})$ by
\[
\tau=\sqrt{1-\nu^2}\tilde{\tau}+b\tilde{\phi}, \quad  \phi=c\tilde{\phi},
\]
where
\begin{eqnarray}
\label{bc}
c&=&-\sqrt{k}\frac{2r_{2}}{a_4(r_2-r_1)(r_2-r_3)(r_2-r_4)}\\
b&=&\sqrt{k}\frac{(1-\nu^2)({r_2}^3+r_1r_3r_4)+2\nu r_2(r_1r_2+r_1r_3+r_1r_4+r_2r_3+r_2r_4+r_3r_4)}{(r_1-r_2)(r_2-r_3)(r_2-r_4)}\nonumber
\end{eqnarray}
so that $\p/\p\tilde{\tau}$ and $\p/\p\tilde{\phi}$ generate the asymptotic Euclidean time translation and rotation respectively and moreover $\p/\p\tilde{\tau}$ agrees with the Killing vector $K$ we used to compute
the patching matrix (\ref{patch_final}). The canonical coordinates $(\tilde{z}, \tilde{r})$
on the space of orbits $M/T^2$ are related to our $(z, r)$ in (\ref{2kf}) by 
\[
z=\gamma^{-1}\tilde{z}, \quad r=\gamma^{-1}\tilde{r}, \quad\mbox{where}\quad \gamma=c\sqrt{1-\nu^2}.
\]
Making the substitution for $z$ in the patching matrix (\ref{patch_final}) removes the dependence
of $a_4$ but introduces the dependence of the scaling parameter $k$ so that the new parameters in the patching matrix are $(k, r_1, r_2, r_3, r_4, \nu)$. To maintain the monic form in $\tilde{z}$
of the cubics $C, C_1, C_2$ in (\ref{patch_final}), the quadratic, linear and $0$th
order terms of $C, C_1, C_2, Q$ scale
with $\gamma, \gamma^2, \gamma^3$ respectively.
This results in scalings of $(m, n)$ in (\ref{massnut})
\be
\label{KLmn}
\tilde{m}=\gamma m, \quad \tilde{n}=\gamma n.
\ee
In particular $\tilde{n}$ is the NUT parameter found by Chen--Teo in \cite{CT2} (compare their formula (3.10)).
\subsection{The Chen--Teo instanton}
\label{subinst}
The expression for the patching matrix simplifies considerably in the case when (\ref{CTmetrics})  is asymptotically flat and regular on the four--manifold
$M=\CP^2\setminus S^1$.  It was shown by Chen and Teo \cite{CT1, CT2}, that
the subclass of metrics (\ref{CTmetrics})  with these properties is parametrised
by $(k, s)$ with 
\[
r_1=\frac{4s^2(1-s)}{1-2s+2s^2},   \quad r_2=-1, \quad r_3=\frac{1-2s}{s(1-2s+2s^2) } , \quad r_4=\infty
\quad \mbox{where}\quad \nu=-2 s^2
\]
and $s\in(1/2, \sqrt{2}/2)$.  We find that the patching matrix is given by (\ref{patch_final}) with
\begin{eqnarray*}
Q&=& \frac{\left(1-s \right) \left(-1+2 s \right) \left(2 s^{2}-1\right)^{3} \left(2 s^{2}-s +1\right)}{2 \left(2 s^{2}+1\right) \left(2 s^{2}-2 s +1\right)^{4}}\Big(z+\frac{s}{2 s^{2}-s +1} \Big)\\ 
C&=&\Big(z+\frac{1}{2}\Big)\Big(z -\frac{2 s^{2} \left(-1+s \right)}{2 s^{2}-2 s +1} \Big)
\Big(z+\frac{-1+2 s}{2 s \left(2 s^{2}-2 s +1\right)}\Big)\\
C_1&=& z^3+c_{12} z^2+c_{11} z+
c_{10}\\
C_2&=&- z^3+c_{22} z^2+c_{21} z+
c_{20}
\end{eqnarray*}
with
\begin{eqnarray*}
c _{10}&=&-\frac{\left(-1+2 s \right) \left(-1+s \right) \left(4 s^{5}-4 s^{4}+2 s^{3}-4 s^{2}+4 s -1\right)}{\left(2 s^{2}+1\right) \left(2 s^{2}-2 s +1\right)^{4}}\\
c_{11}&=&\frac{32 s^{12}-128 s^{11}+240 s^{10}-384 s^{9}+616 s^{8}-768 s^{7}+684 s^{6}-448 s^{5}+224 s^{4}-88 s^{3}+30 s^{2}-8 s +1}{4 s^{2} \left(2 s^{2}+1\right) \left(2 s^{2}-2 s +1\right)^{4}} \\
c_{12}&=& \frac{4 s^{5}-8 s^{4}+8 s^{3}-6 s^{2}+4 s -1}{s \left(2 s^{2}-2 s +1\right)^{2}} \\
c_{20}&=& -\frac{2 \left(8 s^{5}-16 s^{4}+8 s^{3}-2 s^{2}+2 s -1\right) s^{3} \left(-1+2 s \right) \left(-1+s \right)}{\left(2 s^{2}+1\right) \left(2 s^{2}-2 s +1\right)^{4}} \\
c_{21}&=& -\frac{128 s^{12}-512 s^{11}+960 s^{10}-1408 s^{9}+1792 s^{8}-1792 s^{7}+1368 s^{6}-768 s^{5}+308 s^{4}-96 s^{3}+30 s^{2}-8 s +1}{4 \left(2 s^{2}+1\right) \left(2 s^{2}-2 s +1\right)^{4}}  \\
c_{22}&=&\frac{8 s^{5}-16 s^{4}+12 s^{3}-8 s^{2}+4 s -1}{\left(2 s^{2}-2 s +1\right)^{2}}.
\end{eqnarray*}
The asymptotic charges (\ref{massnut}) are
\be
m=\frac{(2s^2-1)(2s^2+1)(s-1)(2s-1)}{4(2s^2-2s+1)}, \quad n=0.
\ee
The vanishing of the NUT parameter $n$ ensures asymptotic flatness,
while the mass parameter $m$ is positive
in the allowed range of $s$. 
The constant $c$ in (\ref{bc}) is now given by $2\sqrt{k}r_2(r_2-r_1)^{-1}(r_2-r_3)^{-1}$ so that 
the scaled mass parameter $(\ref{KLmn})$ 
\[
\tilde{m}=\sqrt{k}\frac{(1+2s^2)^2}{2\sqrt{1-4s^4}}
\]
agrees with the Kunduri--Lucietti mass found in \cite{Lucietti} (see their formula (169)).
\section{An ansatz for the patching matrix}
\label{section3}
Conversely, let us consider a general symmetric patching matrix $P=P(z)$ with rational coefficients and three simple poles, and
such that $\lim_{z=\infty} P=\mbox{diag}(1, -1)$. The most general form of such matrix is
(\ref{patch_final}), with (\ref{coeffC})
but with the cubics $C, C_1, C_2$ and quadratic $Q$ having general coefficients.
Thus $P$ depends on $12$ parameters. The condition
\be
\label{detP}
\det{P}=-1
\ee
gives six conditions on the parameters
\begin{eqnarray*}
0&=&2c_2-c_{12}+c_{22}\\
0&=&c_{12}c_{22}+{c_2}^2-{q_2}^2+2c_1-c_{11}+c_{21}\\
0&=&2c_1c_2+c_{11}c_{22}+c_{12}c_{21}-2q_1q_2+2c_0-c_{10}+c_{20}\\
0&=&2c_0c_2+{c_1}^2+c_{10}c_{22}+c_{11}c_{21}+c_{12}c_{20}-2q_0q_2-{q_1}^2\\
0&=&2c_0c_1+c_{10}c_{21}+c_{11}c_{20}-2q_0q_1\\
0&=&{c_0}^2+c_{10}c_{20}-{q_0}^2.
\end{eqnarray*}
The first of these equations is solved by setting
\[
c_{12}=2m+c_2, \quad c_{22}=2m-c_2,
\]
where the constant $m$ can be identified with the mass
in the asymptotic form (\ref{asymptoticP}). Four of the remaining five equations can be solved for  $(c_0, c_1, c_{10}, c_{20})$, and
the solutions are unique (as the equations are linear). The final  quadratic for $q_0$ can also be solved, but this introduces  branching. Now $(c_{10}, c_{12}, c_{20},
c_{22}, c_0, c_1, q_0)$ have been determined in terms of six constants 
\be
\label{sixconstants}
c_{11}, \quad c_{21},\quad c_2,\quad q_1, \quad q_2,
\quad m.
\ee
We go back to (\ref{thelongest}) and (\ref{massnut})
and consider the six resulting equations for the Chen--Teo parameters $(a_4, r_1, r_2, r_3, r_4, \nu)$ (the same argument applies if we use the parameters $(k, r_1, r_2, r_3, r_4, \nu)$ resulting from the scalings  (\ref{KLmn})). If
we could invert the equations, and thus show that these parameters can be determined
in terms of the constants (\ref{sixconstants}), then we could deduce that the patching matrices with three
simple poles and these asymptotics result precisely from the Chen--Teo metrics. A solution to the six equations is the intersection of six hyper-surfaces. Therefore it is enough to show that the six normal vector fields are linearly independent. Computing the gradients of (\ref{sixconstants}) and  using  (\ref{thelongest}) and (\ref{massnut}), we find that they are in fact linearly {\em dependent} everywhere, and the resulting $6$ by $6$ matrix $\mathcal{M}$ of coefficients has rank $5$. There is a residual
freedom in $P$ which preserves its symmetry and the asymptotic form $\mbox{diag}(1, -1)$. It is given by
\[
P\rightarrow LPL^{T}, \quad\mbox{where}\quad L= \begin{pmatrix}
\cosh{t} & \sinh{t}   \\
\sinh{t} & \cosh{t}
\end{pmatrix}  \in SO(1, 1)
\]
and $t$ is a constant. This results in
\begin{eqnarray*}
C_1(t) &=& C_1 \cosh{t}^2 +C_2\sinh(t)^2 +Q\sinh{(2t)}\\
C_2(t)&=&  C_1 \sinh{t}^2 +C_2\cosh(t)^2 +Q\sinh{(2t)} \\
Q(t)&=& Q (\cosh{(t)}^2+\sinh{(t)}^2)+\frac{1}{2}(C_1+C_2)\sinh{(2t)},
\end{eqnarray*}
and leaves the cubics $C$ and $C_1-C_2$ invariant. It is however the case that the rank of the transformed matrix $\mathcal{M}(t)$ is still $5$. Therefore there is  one invariant syzygy between the parameters (\ref{sixconstants}) coming from the Chen--Teo metrics.  Using the combined results of \cite{AA} and \cite{BG}
suggests that the ansatz (\ref{patch_final}) contains some toric--Einstein metrics which are not Hermitian or not regular in that their singularities are worse than conical.

%{\em [21.4.2024] I have now done it, and it does not cure the problem. The calculation can be done on %Maple for a general Lorentz transformation, or at an infinitesimal level. If the Lorentz parameter is $t$, then %to the first order 
%\[
%C_1\rightarrow C_1+2tQ, \quad C_2\rightarrow C_2+2tQ, \quad Q\rightarrow Q+t(C_1+C_2),
%\]
%with the resulting transformation of coefficients. The relevant Jacobian is still zero in $t$, as we are unable to %pick six independent coeffs.}

%\item We may have picked {\em wrong} six parameters to parametrise $C, C_1, C_2, Q$.

%{\em [21.4.2024] I have tried other parametrisations}

%\item Chen--Teo is not the most general solution.
%\end{enumerate}
\appendix
\section{Twistor correspondence for the Yang equation}
\renewcommand{\theequation}{A.\arabic{equation}}
In this Appendix we shall summarise the Ward twistor correspondence \cite{W1} for anti--self--dual Yang--Mills (ASDYM) equations in a gauge where they are equivalent to a single non--linear matrix PDE. We shall follow the notation and conventions of \cite{D}.
\subsection{The Yang equation} Let $(W, Z, \tw, \tz)$ be double null coordinates on the complexified Minkowski space $M_{\C}=\C^4$ with the holomorphic metric
$ds^2$ and the volume form $\mbox{vol}$ given by
\be
\label{metric}
ds^2=2(dZ d\tz-dWd\tw), \quad \mbox{vol}=dW\wedge d\tw \wedge dZ \wedge d\tz.
\ee
Let $A$ be an $\mathfrak{sl}(2)$--valued connection one--form with the curvature $F=dA+A\wedge A$ and a covariant derivative $D=d+A$. The ASDYM condition $F=-\star F$ takes the form
\be
\label{ASDYM}
F_{WZ}=0, \quad F_{\tw\tz}=0,\quad  F_{W\tw}-F_{Z\tz}=0.
\ee
These equations arise as the compatibility conditions $[L, M]=0$ for an over--determined
linear system of PDEs 
\be
\label{lax}
L\Psi=0, \quad M\Psi=0, \quad\mbox{where}\quad L=D_{\tz}-\lambda D_{W}, \quad M=D_{\tw}-\lambda D_{Z}.
\ee
Here $\lambda\in\CP^1$ is the spectral parameter, and $\Psi=\Psi(W, Z, \tw, \tz, \lambda)$ takes its values in the group $SL(2, \C)$ (other gauge groups are also possible).

The first equation in (\ref{ASDYM}) implies the existence of a gauge such that $A_W=A_Z=0$. The second equation then implies the existence of an $SL(2, \C)$ valued function $J=J(W, Z, \tw, \tz)$ such that
\be
\label{specialgauge}
A=J^{-1}\p_{\tw} J d\tw+J^{-1}\p_{\tz} J d\tz.
\ee
% Define two $SL(2, \C)$ valued functions $H, \widetilde{H}$ on $M_{\C}$ by
%\[
%\Psi(w, \zeta, \tw, \tz, \lambda=0)=H^{-1}, \quad \Psi(w, \zeta, \tw, \tz, \lambda=\infty)=\widetilde{H}^{-1}.
%#%\]
%Then $J\equiv \widetilde{H}^{-1}H$ satisfies the Yang--equation
Finally the last equation in (\ref{ASDYM}) becomes the so--called Yang equation
\be
\label{yang}
 \p_Z(J^{-1}\p_{\tz} J)-  \p_W(J^{-1}\p_{\tw} J)=0.
\ee
Therefore the Yang equation is equivalent to ASDYM together with a particular choice of gauge.
\subsection{Twistor correspondence}
\label{sectionA2}
The ASDYM equations (\ref{ASDYM}) are equivalent to the flatness of the connection $A$ on self--dual
totally null two--planes ($\alpha$--planes in twistor terminology). All such planes in $M_{\C}$ are
of the form
\be
\label{incidence}
\alpha=W+\lambda\tz, \quad \beta= Z+\lambda\tw
\ee
and the space of these planes is the projective twistor space $PT\equiv \CP^3\setminus\CP^1$ with affine coordinates $(\alpha, \beta, \lambda)$ covering the open set where $\lambda
\neq \infty$. The vector space of parallel sections  of the Yang--Mills
bundle  ${\mathcal E}\rightarrow M_{\C}$
on  each $\alpha$--plane defines a fibre of a holomorphic rank--2 vector bundle $E\rightarrow PT$. For any fixed point 
$p\in M_{\C}$ the equations (\ref{incidence}) define a rational curve
$L_p\cong \CP^1$ in the twistor space. The sections
of $\mathcal{E}$ constant on all $\alpha$--planes through $p$ can be compared at $p$, and thus the
twistor bundle $E$ is trivial on all lines (\ref{incidence}).

Conversely, let $E\longrightarrow PT$ be a rank--2 vector bundle trivial on lines. Cover $PT$ with two open sets: $U$ where $\lambda\neq \infty$, and $\widetilde{U}$
where $\lambda \neq 0$. Let $P_{U\widetilde{U}}=P(\alpha, \beta, \lambda)$ be the patching matrix
of the bundle which is holomorphic on $U\cap \widetilde{U}$. Restrict $P$ to a twistor line (\ref{incidence}) where the bundle is trivial so that a Riemann--Hilbert splitting 
\be
\label{Apatching}
P_{U\widetilde{U}}=P_U{{P}_{\widetilde{U}}}^{-1}
\ee
can be performed with $P_U$ and $P_{\widetilde{U}}$
holomorphic and non--singular in $U$ and $\widetilde{U}$ respectively. The patching matrix $P_{U\widetilde{U}}$ is constant
over the $\alpha$--planes spanned by 
$l=\p_{\tz}-\lambda \p_{W}$,  and $m=\p_{\tw}-\lambda \p_{Z}$. Therefore the quantities
$
{P_{U}}^{-1} l (P_U)={P_{\widetilde{U}}}^{-1} l (P_{\widetilde{U}}),  \quad {P_{U}}^{-1} m (P_U)={P_{\widetilde{U}}}^{-1} m (P_{\widetilde{U}})
$
are global on $\CP^1$, and by the application
of the Liouville theorem there exist
$\mathfrak{sl}(2, \C)$ valued one--form $A$
on the complexified Minkowski space such that
\be
\label{PUH}
P_{U}^{-1} l (P_U)=  {P_{\widetilde{U}}}^{-1} l (P_{\widetilde{U}})=  A_{\tz}-\lambda A_W, \quad
P_{U}^{-1} m (P_U)=    {P_{\widetilde{U}}}^{-1} m (P_{\widetilde{U}})= A_{\tw}-\lambda A_{Z}.
\ee
Setting $H=P_U(\lambda=0), \widetilde{H}=P_{\widetilde{U}}(\lambda=\infty)$ then gives
\[
A=H^{-1}\p_Z H\; dZ+H^{-1}\p_W H\;dW+\widetilde{H}^{-1}\p_{\tz} \widetilde{H}\; d\tz+
\widetilde{H}^{-1}\p_{\tw} \widetilde{H}\; d\tw
\]
which is gauge--equivalent to (\ref{specialgauge}) with $J=\widetilde{H}H^{-1}$. Applying $l$ and $m$ to the first and the second relation in (\ref{PUH}) and commuting derivatives then implies that $A$ is a solution of ASDYM, or equivalently that $J$ satisfies the Yang equation.
\subsection{Symmetry reduction}
We shall use the cylindrical polar coordinates
\[
Z=t+z, \quad \tz=t-z, \quad W=re^{i\theta}, \quad
\tw=re^{-i\theta}
\]
so that
\[
ds^2=2(dt^2-dz^2-dr^2-r^2d\theta^2),
\]
and assume that the gauge potential is invariant under a 2--dimensional Abelian group generated
by a translation $\p_t$ and a rotation $\p_{\theta}$. We chose  $J=J(r, z)$ in which case the Yang equation (\ref{yang}) reduces to equation (\ref{Yang}) locally characterising the toric Ricci--flat four--manifolds.
The Killing vectors ${\mathcal{K}_1}=\p_t$ and ${\mathcal{K}_2}= \p_{\theta}$ on $M_{\C}$ give rise to holomorphic vector fields $({\hat{\mathcal{K}}_1}, {\hat{\mathcal{K}}_2})$
on $PT$. These vector fields have zeros, so the reduced twistor space defined to be the corresponding quotient
of $PT$ is a one--dimensional non--Hausdorff complex manifold \cite{F, FW, WM}. Rather than taking this quotient we can consider the holomorphic vector bundle $E\rightarrow PT$ as defined in \S\ref{sectionA2}
such that the patching matrix (\ref{Apatching}) depends only on one complex variable  $\gamma$ constant along $({\hat{\mathcal{K}}_1}, {\hat{\mathcal{K}}_2})$ . The incidence relation (\ref{incidence}) then reduces to
\[
\gamma=z+\frac{1}{2}r\Big(\lambda-\frac{1}{\lambda}\Big).
\]
Therefore knowing the patching matrix at a rod corresponding to a fixed value of $r$, as in formula (\ref{patch_final}) suffices (note that $iP$ in this formula belongs to $SL(2, \C)$), by analytic continuation, to determine it completely on an intersection of open sets covering $PT$. If two solutions of the Yang equation (\ref{Yang}) are related by the 
B\"acklund transformation (\ref{Jprime}), then the corresponding patching matrices are related by a conjugation
\[
P\rightarrow \begin{pmatrix}
1 & 0   \\
\gamma^{-1} & 1
\end{pmatrix}^{-1} P \begin{pmatrix}
1 & 0   \\
\gamma^{-1} & 1
\end{pmatrix}.
\]
\newpage
\section{Coefficients of the patching matrix}
\renewcommand{\theequation}{B.\arabic{equation}}
Coefficients of the cubics and the quadratic  (\ref{coeffC}) 
\footnote{The expressions
for $q_0, q_1$ are long and can be obtained from a  MAPLE worksheet on request.}
\begin{eqnarray}
\label{thelongest}
c_0&=& \frac{{a_4}^3}{8}\left( r_{1} r_{3}+ r_{2} r_{4}\right) \left( r_{1} r_{2}+ r_{3} r_{4}\right) \left( r_{1} r_{4}+ r_{2} r_{3}\right) \nonumber\\
c_1&=& \frac{{a_4}^2}{4}  \Big( \left(r_{1} r_{3}+r_{2} r_{4}\right) \left(r_{1} r_{2}+r_{3} r_{4}\right)+
\left(r_{1} r_{2}+ r_{1} r_{3}+ r_{2} r_{4}+r_{3} r_{4}\right) \left( r_{1} r_{4}+ r_{2} r_{3}\right)\Big)\\
c_2&=&\frac{a_{4}}{2}  \left(r_{1} r_{2}+r_{1} r_{3}+r_{1} r_{4}+r_{2} r_{3}+r_{2} r_{4}+r_{3} r_{4}\right)  \nonumber\\
c_{10}&=& \frac{1}{\left(4 \nu^{2}-4\right) r_{2}}\Big(
a_{4}^{3} \left(\nu  \left(r_{3}r_1+r_{4} r_{1}+r_{3} r_{4}\right) r_{2}^{2}-r_{1} r_{3} r_{4} \left(\nu -1\right) r_{2}-r_{1} r_{3} r_{4} \left(r_{1}+r_{3}+r_{4}\right)\right) \Big)\nonumber\\
 &&\Big(\left(-\nu  r_{2}^{3}+\nu  \left(r_{1}+r_{3}+r_{4}\right) r_{2}^{2}+\left(r_{3}r_1+r_{4}r_{1}+r_{3} r_{4}\right) r_{2}-r_{1} r_{3} r_{4}\right)\Big) 
 \nonumber\\
c_{11}&=&\frac{1} {\left(4 -4\nu^2 \right) r_{2}}\Big(
a_{4}^{2} (\nu^{2} r_{2}^{4} r_{1}^{2}+4 \nu^{2} r_{1} r_{2}^{4} r_{3}+4 \nu^{2} r_{1} r_{2}^{4} r_{4}-2 r_{1} r_{3} r_{4} r_{2}^{3} \nu^{2}-\nu^{2} r_{2}^{6}+\nu^{2} r_{2}^{4} r_{3}^{2}+4 \nu^{2} r_{2}^{4} r_{3} r_{4}\nonumber\\
&&
+\nu^{2} r_{2}^{4} r_{4}^{2}-4 r_{1}^{2} r_{3} r_{4} r_{2}^{2} \nu +4 \nu  r_{1} r_{2}^{4} r_{3}+
4 \nu  r_{1} r_{2}^{4} r_{4}-4 r_{1} r_{3}^{2} r_{4} r_{2}^{2} \nu -4 r_{1} r_{3} r_{4}^{2} r_{2}^{2} \nu +4 \nu  r_{2}^{4} r_{3} r_{4}-r_{1}^{2} r_{2}^{2} r_{3}^{2}\nonumber\\
&&
-4 r_{1}^{2} r_{3} r_{4} r_{2}^{2}
-r_{1}^{2} r_{2}^{2} r_{4}^{2}
+r_{1}^{2} r_{3}^{2} r_{4}^{2}+2 r_{1} r_{3} r_{4} r_{2}^{3}-4 r_{1} r_{3}^{2} r_{4} r_{2}^{2}-4 r_{1} r_{3} r_{4}^{2} r_{2}^{2}-r_{2}^{2} r_{3}^{2} r_{4}^{2})\Big)
\nonumber\\
c_{12}&=&  \frac{a_{4} \left(\nu  r_{2}^{3}+\nu^{2} \left(r_{1}+r_{3}+r_{4}\right) r_{2}^{2}
-\left(r_{3}r_1+r_{4} r_{1}+r_{3} r_{4}\right) r_{2}-\nu  r_{1} r_{3} r_{4}\right)}{r_{2} \left(\nu^{2}-1\right)} \nonumber
\end{eqnarray}
\begin{eqnarray}
c_{20}&=&
\frac{1} {\left(4 -4\nu^2 \right) r_{2}}
\Big(
{a_{4}}^{3} \left(\left(-r_{3}r_1-r_{4} r_{1}-r_{3} r_{4}\right) r_{2}^{2}-r_{1} r_{3} r_{4} \left(\nu -1\right) r_{2}+\nu  r_{1} r_{3} r_{4} \left(r_{1}+r_{3}+r_{4}\right)\right)\Big)\nonumber\\
&& \Big(-r_{2}^{3}+\left(r_{1}+r_{3}+r_{4}\right) r_{2}^{2}+\nu  \left(r_{3}r_1+r_{4} r_{1}+r_{3} r_{4}\right) r_{2}-\nu  r_{1} r_{3} r_{4}\Big)
\nonumber\\
c_{21}&=& -\frac{1} {\left(4 -4\nu^2 \right) r_{2}}\Big(
a_{4}^{2} (r_{1}^{2} r_{2}^{2} r_{3}^{2} \nu^{2}+4 r_{1}^{2} r_{3} r_{4} r_{2}^{2} \nu^{2}
+r_{1}^{2} r_{2}^{2} r_{4}^{2} \nu^{2}-r_{1}^{2} r_{3}^{2} r_{4}^{2} \nu^{2}-2 r_{1} r_{3} r_{4} r_{2}^{3} \nu^{2}+4 r_{1} r_{3}^{2} r_{4} r_{2}^{2} \nu^{2}\nonumber\\
&&+4 r_{1} r_{3} r_{4}^{2} r_{2}^{2} \nu^{2}+r_{2}^{2} r_{3}^{2} r_{4}^{2} \nu^{2}+4 r_{1}^{2} r_{3} r_{4} r_{2}^{2} \nu -4 \nu  r_{1} r_{2}^{4} r_{3}-4 \nu  r_{1} r_{2}^{4} r_{4}+4 r_{1} r_{3}^{2} r_{4} r_{2}^{2} \nu +4 r_{1} r_{3} r_{4}^{2} r_{2}^{2} \nu \nonumber\\
&&-4 \nu  r_{2}^{4} r_{3} r_{4}-r_{2}^{4} r_{1}^{2}-4 r_{1} r_{2}^{4} r_{3}-4 r_{1} r_{2}^{4} r_{4}+2 r_{1} r_{3} r_{4} r_{2}^{3}+r_{2}^{6}-r_{2}^{4} r_{3}^{2}-4 r_{2}^{4} r_{3} r_{4}-r_{2}^{4} r_{4}^{2})\Big)\nonumber\\
c_{22}&=&\frac{a_{4} \left(\nu  r_{2}^{3}+\left(r_{1}+r_{3}+r_{4}\right) r_{2}^{2}-\left(r_{3}r_1+
r_{4} r_{1}+r_{3} r_{4}\right) \nu^{2} r_{2}-\nu  r_{1} r_{3} r_{4}\right)}{r_{2} \left(\nu^{2}-1\right)}  \nonumber\\
%q_0&=&\nonumber\\
%q_1&=&\nonumber\\
q_2&=& \frac{\left(-2 \nu  \left(r_{1}+r_{3}+r_{4}\right) r_{2}^{2}+2 \nu  \left(r_{3}r_1+r_{4}r_{1}+r_{3} r_{4}\right) r_{2}+ \left(\nu^{2}+1\right)( r_4r_{3} r_{1} -{r_2}^3) \right) a_{4}}{\left(2 \nu^{2}-2\right) r_{2}}.
\end{eqnarray}

\end{document}